\documentclass[referee]{aa} 

\usepackage{graphicx}
\usepackage{txfonts}
\begin{document}

   \title{TOPoS} 

   \subtitle{V: Abundance ratios in a sample of very metal-poor turn-off stars}
   \titlerunning{ $\alpha$ and neutron$-$capture elements abundance in turn-off metal-poor stars}      
  \author{P. Fran\c cois
          \inst{1,2}
          \and
          E. Caffau\inst{3}
            \and
          P. Bonifacio \inst{3}    
          \and
          M. Spite  \inst{3}  
          \and 
          F. Spite  \inst{3}  
          \and
           R. Cayrel  \inst{1} 
          \and
          N. Christlieb  \inst{4} 
          \and
          A. J. Gallagher  \inst{5} 
          \and 
          R. Klessen   \inst{6} 
          \and
          A. Koch     \inst{7} 
          \and
          H.-G. Ludwig  \inst{4} 
          \and
          L. Monaco \inst{8} 
          \and
            B. Plez \inst{9} 
          \and        
          M. Steffen \inst{10}
          \and
          S. Zaggia \inst{11}   
          \thanks{Based on observations collected at the European Organisation for Astronomical Research in the Southern Hemisphere under  ESO programme ID 189.D-0165}
        }
   \institute{GEPI, Observatoire de Paris,  Universit\'e PSL, CNRS,  61 Avenue de l'Observatoire, 75014 Paris, France \\
              \email{patrick.francois@obspm.fr} 
         \and
             Universit\' e de Picardie Jules Verne, 33 rue St Leu, Amiens, France 
             \and 
             GEPI, Observatoire de Paris, Universit\'e PSL, CNRS, Place Jules Janssen, 92190 Meudon, France  
             \and
             Zentrum f\"ur Astronomie der Universit\"at Heidelberg, Landessternwarte, K\"onigstuhl 12, 69117, Heidelberg, Germany
             \and
             Max Planck Institute for Astronomy, K\"onigstuhl 17, 69117, Heidelberg, Germany
             \and
             Zentrum f\"ur Astronomie der Universit\"at Heidelberg, Institut f\"ur Theoretische Astrophysik, Albert-Ueberle-Strasse 2, 69120 Heidelberg, Germany
             \and
             Zentrum f\"ur Astronomie der Universit\"at Heidelberg, Astronomisches Recheninstitut, M\"onchhofstr. 12, 69120 Heidelberg, Germany
             \and
             Departamento de Ciencias Fisicas, Universidad Andres Bello, Fernandez Concha 700, Las Condes, Santiago, Chile
             \and
             Laboratoire Univers et Particules de Montpellier, LUPM, Universit\'e de Montpellier, CNRS, 34095, Montpellier cedex 5, France
             \and
             Leibniz-Institut f\"ur Astrophysik Potsdam (AIP), An der Sternwarte 16, 14482, Potsdam, Germany
             \and
             Istituto Nazionale di Astrofisica, Osservatorio Astronomico di Padova, Vicolo dell'Osservatorio 5, 35122, Padova, Italy
             }

   \date{Received ; accepted }

 
  \abstract
   { Extremely metal$-$poor stars are keys to understand the early  evolution of our Galaxy. The  ESO large programme TOPoS has been tailored to analyse a new set of metal$-$poor turn$-$off stars, whereas most of the previously known extremely metal$-$poor stars are giant stars. }
   { Sixty five  turn$-$off stars (preselected from SDSS spectra) have been observed with the X$-$Shooter spectrograph at the ESO VLT Unit Telescope 2, to derive accurate and detailed abundances of magnesium, silicon, calcium, iron, strontium and barium.}
   { We analysed medium$-$resolution spectra ( R $\simeq$ 10 000) obtained with the  ESO X$-$Shooter spectrograph and computed the abundances of several $\alpha$ and neutron$-$capture elements  using standard one$-$dimensional local thermodynamic equilibrium (1D LTE) model atmospheres. 
  }
   { Our results confirms the super-solar [Mg/Fe] and  [Ca/Fe] ratios in metal$-$poor turn$-$off stars  as observed in  metal$-$poor giant stars. We found a significant spread of the
   [$\alpha$/Fe] ratios with several stars showing
 sub$-$solar [Ca/Fe] ratios.  We could measure the abundance
   of strontium in 12 stars of the sample, leading to abundance ratios  [Sr/Fe] around the Solar value. We detected barium in two stars of the sample.   
   One of the stars (SDSS J114424$-$004658) shows both  very high 
 [Ba/Fe] and [Sr/Fe] abundance ratios (>~1~dex).
   }
   {}
   
 \authorrunning{P. Fran\c{c}ois et al.}
\maketitle
%

\section{Introduction}

The study of the chemical composition of metal$-$poor stars is one of the most important tools to understand and constrain the models for the early evolution of 
our Galaxy    \citep[see][and reference therein]{beers2005, frebel2015}.  In particular,  detailed abundance ratios of the stars of the lowest [Fe/H] in our Galaxy contain  information on the nature  of  the first stars  which enriched 
 the  gas which subsequently formed the most metal$-$poor stars we are observing. 
 TOPoS (Turn$-$Off PrimOrdial Stars) is a survey based on the
ESO/VLT Large Programme 189.D-0165. The observation programme
spanned four ESO semesters (Period 89 to period 92), from April 2012 to
March 2014, for a total of 120 h with X-Shooter, and 30 h with UVES. 
 The first paper of the series \citep{caffau2013a} contains  the details of the TOPoS project. 
The main objectives of the survey are the following : 
\begin{itemize}
\item{} search for the extremely metal-poor (EMP) stars.
\item{} understand the formation of low$-$mass stars in a metal$-$poor environment. 
\item{} use the detailed chemical composition of the most metal$-$poor star to constrain the masses of the first Pop. III stars.
\item{} determine the Lithium abundance in these EMP stars.
\item{} derive the fraction the fraction of C$-$enhanced EMP stars with respect normal EMP stars. 
\end{itemize}
More details about these objectives can be found in \citet{caffau2013a}.

  In the present paper, we focus on the determination of the chemical composition ($\alpha$ and neutron$-$capture elements) of the stars of the TOPoS sample. 
Eighty$-$four metal$-$poor candidates have been  followed-up
with X$-$Shooter in total, and the most interesting stars, seven
stars, have been  followed$-$up with UVES. The analysis of the  UVES targets  can be found in \citet{bonifacio2018}.
The results concerning the first sample of 19 stars observed with X$-$Shooter are presented in \citet{caffau2013b}.
In this article, we report on the  abundances of magnesium, silicon, calcium, strontium and barium for the remaining sample of 65 stars
  which have been observed with X$-$Shooter.

\section{Observations}

The observations were performed in Service Mode with Kueyen
(VLT$-$UT2) and the high-efficiency spectrograph X$-$Shooter
\citep{dodo2006, vernet2011}. The X-Shooter spectra
range from 300\,nm to 2400\,nm and are gathered by three detectors.
The observations have been performed in staring mode
with 1$\times$1 binning and the integral field unit (IFU), which reimages
an input field of  4.0$\times$1.8\arcsec into a pseudo-slit of 12.0$\times$0.6\arcsec
\citep{guinouard2006}. As no spatial information was available
for our point$-$source targets, we used the IFU as a slicer with three  0.6\arcsec
slices. This corresponds to a resolving power of R = 7900 in
the ultra$-$violet arm (UVB) and R = 12 600 in the visible arm
(VIS). Although the stellar light is divided into three arms by X$-$Shooter (UVB, VIS and BIR), 
we only analysed the UVB and VIS spectra in this paper. The stars we
observed are rather faint and have most of their detected flux in the blue  and visible parts
of the spectrum, so that the signal-to-noise ratio (S/N) of the
infra-red spectra is too low to allow for any meaningful analysis to be conducted.

The spectra
were reduced using the X$-$Shooter pipeline \citep{goldo2006},
which performs the bias and background subtraction, cosmic ray$-$hit 
removal \citep{vandok2001}, sky subtraction \citep{kelson2003},
 flat-fielding, order extraction, and merging. However, the
spectra were not reduced using the IFU pipeline recipes. Each of
the three slices of the spectra were instead reduced separately in
slit mode with a manual localisation of the source and the sky.
This method, which is not implemented in the current pipeline, 
 allowed us to perform the best possible extraction
of the spectra, leading to an efficient cleaning of the remaining
cosmic ray hits, but also to a noticeable improvement in the S/N.
Using the IFU can cause some problems with the sky subtraction, 
because there is only $\pm$1$\arcsec$ on both sides of the object.
In the case of a large gradient in the spectral flux (caused by
emission lines), the modeling of the sky-background signal can
be of poor quality owing to the small number of points used in
the modeling. We experienced these difficulties  for the NIR spectra and the reddest orders of the VIS spectra.   
For this analysis, the lines we measured were located in  the wavelength range 390 - 650 nm.
 
\section{Stellar parameters}

The stellar parameters  shown in Table \ref{stellar_parameters} have  been computed following the method described in detail in  \citet{caffau2013a}. 
Table \ref{stellar_parameters} includes  the coordinates, the G magnitudes from Gaia  DR2 \citep{gaia2016,gaia2018} and the SNR of the spectra.
The effective temperature has been derived from the photometry, using the 
$(g-z)_{0}$ colour and the calibration described in \citet{ludwig2008},  taking into account  the 
reddening according to the \citet{schlegel1998} extinction maps and 
corrected using the techniques described in \citet{bonifacio2000}. As the stars of the sample were selected to have typical TO
colours, an archetypical gravity of $\log g = 4.0$ has been assumed for all stars.  The micro-turbulent velocity has been set to 1.5 km/s. 
The determination of the [Fe/H] abundances
was computed using the code MyGIsFOS  \citep{sbordone2014}. This code is based on the selection of clean Fe lines for which the equivalent width is measured \footnote{ Equivalent widths of the Fe lines are only available at the CDS.}.
 Further details  about the determination of the [Fe/H]  can be found in \citet{caffau2013a}. We do not report the radial velocities of the stars as it will be discussed in detail in a forthcoming paper.

\section{Analysis}

We carried out a classical 1D LTE analysis using 
MARCS model atmospheres  \citep{gustafsson2008}.
The abundances used in the model atmospheres
were solar-scaled with respect to the \citet{grevesse2000} solar abundances,
except for the $\alpha$$-$elements that are enhanced by 0.4\,dex. We corrected the resulting abundances by taking into account the difference
between  \citet{grevesse2000}  and \citet{caffau2011}, \citet{lodders2009}  solar abundances. The final adopted solar abundances in this paper are: A(C)=8.50, A(Ca)=6.33, A(Mg)=7.54, A(Fe)=7.50, A(Si)=7.52, A(Sr)=2.92 and  A(Ba)=2.17.

The abundance analysis was performed using the LTE spectral line analysis code turbospectrum \citep{alvarez1998, plez2012}, which includes continuum scattering in the source function \citep{cayrel2004}. For each available transition, we computed a synthetic spectrum and compared  the synthetic spectrum directly with the observed spectrum.  We used the three lines of the magnesium triplet to derive the magnesium abundance. For silicon, we used the 390.5 nm line. 
 The calcium abundance was measured using the lines at 422.6, 443.4 and 445.5 nm. We did not use the Ca II IR lines which is very likely caused by an incorrect sky subtraction. These lines are also known to be sensitive to departures from  LTE, and are usually too strong for meaningful abundance determination.
For strontium, we used the lines at 407.7  and 421.5 nm. Barium abundances  were determined using the 455.4 nm line. 
 We also evaluated the upper limits for the the carbon abundance  by fitting the CH G$-$band.  Given that we selected our targets from spectra with weak
CH G$-$band,  we do expect to find stars with a high carbon abundance. Indeed, we did not find any star with a high [C/Fe] ratio compatible with the high$-$carbon  band as defined by \citet{spite2013}.  From our measurements, a fraction of the stars  populates the low$-$carbon band of the CEMP stars, but these upper limits are based on spectra with a rather low SNR and may give much higher carbon abundances than the true value.
Indeed, the high [C/Fe] upper limits  we measured are all found in  the hottest stars of the sample,
   a temperature range for which the determination of the carbon abundance is particularly challenging. 
Further analyses with spectra of higher quality would help to give better estimates of the carbon abundance in these stars.

The abundances tabulated in  Table \ref{star_abund} represent the best fit to the data. In some cases only a single transition has been measured providing a single abundance, whereas in others several lines have been measured and an average abundance is presented. The SNR of the reduced spectra 
presented in table \ref{stellar_parameters} may be used to evaluate the quality of the spectra. For the different elements studied in this article, we give the number of absorption lines  that have been used to determine the abundances  in Table \ref{star_abund}.

\section{Errors}

Table \ref{errors} lists the computed errors in the elemental abundances ratios due to typical uncertainties in the stellar parameters. The errors were estimated varying  $T_{\mathrm { eff}}$ by $\pm$ 100~K, $\log~g$  by $\pm$  0.3~dex and $ v_{t}$  by $\pm$ 0.5 dex in the model atmosphere of SDSS~J154746+242953, other stars give similar results. We chose this star because it has a temperature and ${\rm [Fe/H]}$  close to the median value of the ranges for the sample. Moreover,  we could measure the Mg, Si, Ca, Ba and Sr abundances in this star.  The main uncertainty comes from the error in the placement of the continuum when the synthetic line profiles are matched to the observed spectra. This error is of the order of 0.2 to 0.3 dex depending on the species under consideration.  When several lines are available, the typical line$-$to$-$line scatter for a given elements is 0.1 to 0.2 dex.

\section{Results and discussion}

\begin{table}
 \caption{Estimated errors in the element abundance ratios [X/Fe] for the star  SDSS~J154746+242953. The other stars give similar results.} 
\label{errors}
\centering
\begin{tabular}{l r r r }
\hline\hline
  [X/Fe]            &   $\Delta T_{\mathrm {eff}}$ =   & $\Delta$ $\log~g$ =   &$\Delta$  $ v_{t}$ =   \\     
                        &       100~K                 &    0.3~dex        &     0.5~km/s \\  
\hline
Mg &   0.1  &  -0.1 &  -0.15  \\
Ca &   0.15  &  -0.1  &  0.15 \\  
Si &   0.1 &  -0.1  &  -0.15  \\
Sr &  -0.2  &  0.2  &  0.25  \\
Ba &   0.2  &  0.2   &  -0.3  \\
  \hline
  \end{tabular}
   \end{table}

For most of the stars, we could derive the abundance of magnesium, silicon and  calcium thanks to the strong lines of these  species. We could also determine the abundance of strontium and barium for some of the targets. The results are gathered in Table \ref{star_abund}. 


  \begin{figure}
   \centering
 \resizebox{\hsize}{!} {\includegraphics[clip=true]{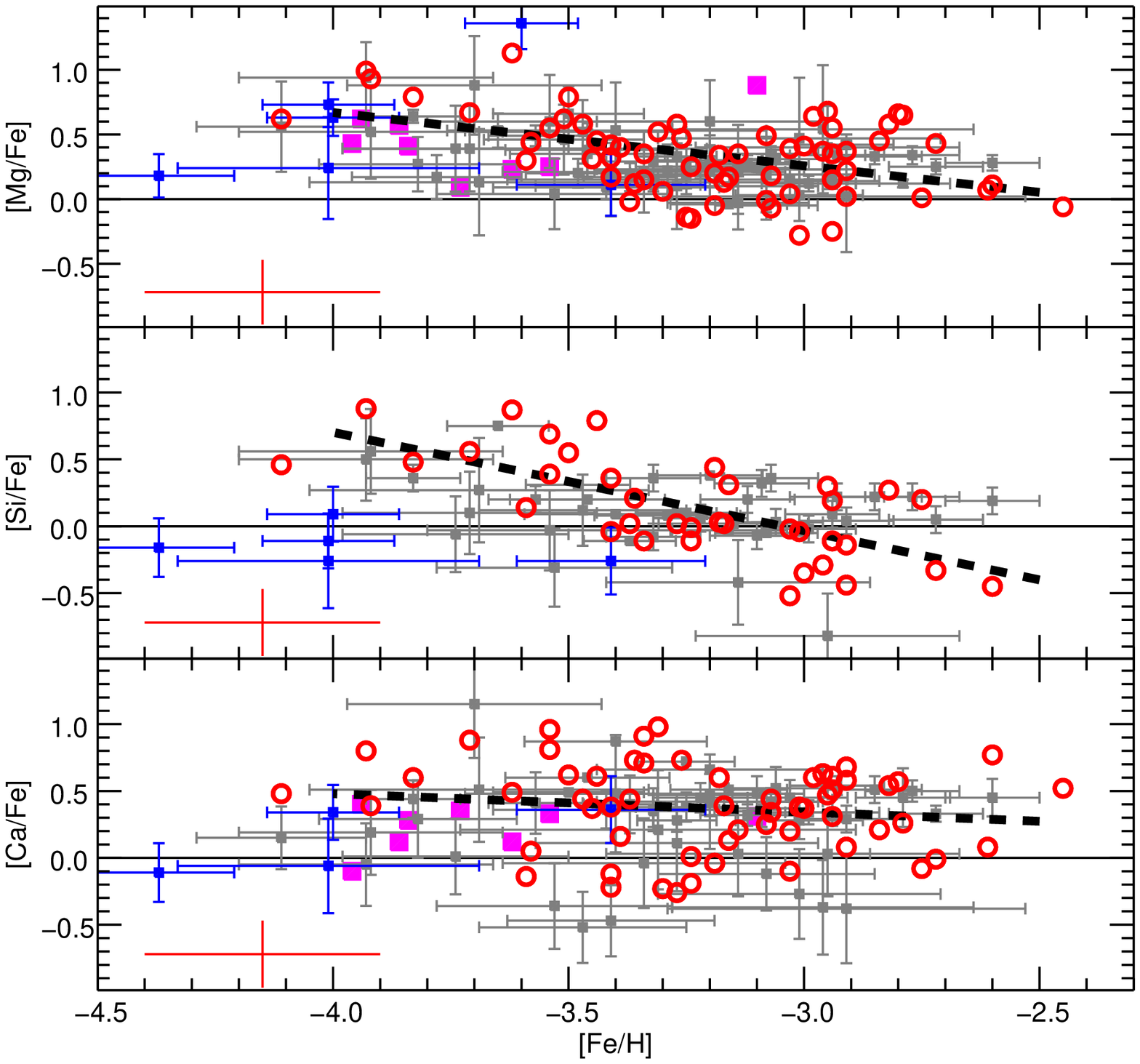}}
   \caption{Abundances in unevolved stars : the ratios [Mg/Fe] , [Si/Fe] and [Ca/Fe] for the program stars (red open circles) compared to those measured  (grey and blue squares) by our group in other extremely metal-poor dwarf stars \citep{bonifacio2009,  bonifacio2012, bonifacio2018,  caffau2013a,  caffau2013b}. Blue (resp. grey) squares represent the stars which have been observed with UVES (resp. Xshooter).  Pink squares represent dwarf stars analysed by \citet{matsuno2017}.  Black dashed lines represent the linear fit to our data. Typical errors are represented in the lower left par of each panel.   }
              \label{alpha}%
    \end{figure}

\subsection{Magnesium, calcium and silicon in turn$-$off stars.}

In Fig.\ref{alpha},  we plot our results as red circles together with previous results from our group  \citep{bonifacio2009,  bonifacio2012, bonifacio2018,  caffau2013a,  caffau2013b}.
We add the results from \citet{matsuno2017}. It is important to note that this figure contains only turn$-$off stars. 
  The results we obtain for the new set of data seem to confirm previous results found for turn$-$off stars, i.e. a slight increase of  the [Mg/Fe] and [Si/Fe] abundance ratios
  at low [Fe/H], and a constant [Ca/Fe]  abundance ratio, although with a larger scatter. The [Ca/Fe] ratios appear to be
 constant, on average, super-solar over the entire range of [Fe/H] values depicted.  However, the results concerning dwarf stars analysed  from UVES spectra by our group (blue symbols)  seem to give lower abundance ratios.   Below [Fe/H] $\simeq$ $-$4.0 dex, the abundance ratios  found by previous studies seem to decrease down to values close to solar. This result should be taken with caution as it is based on few stars.   
  If we consider the full set of results for turn$-$off stars shown in  Fig.\ref{alpha}, the abundance ratios can be interpreted with a constant super-solar   [Mg/Fe], [Si/Fe] and [Ca/Fe] ratios
as found in giant stars \citep{cayrel2004}, although with a larger scatter.

 \subsection{Abundance spread}

  \begin{figure}
   \centering
 \resizebox{\hsize}{!}{\includegraphics[clip=true]{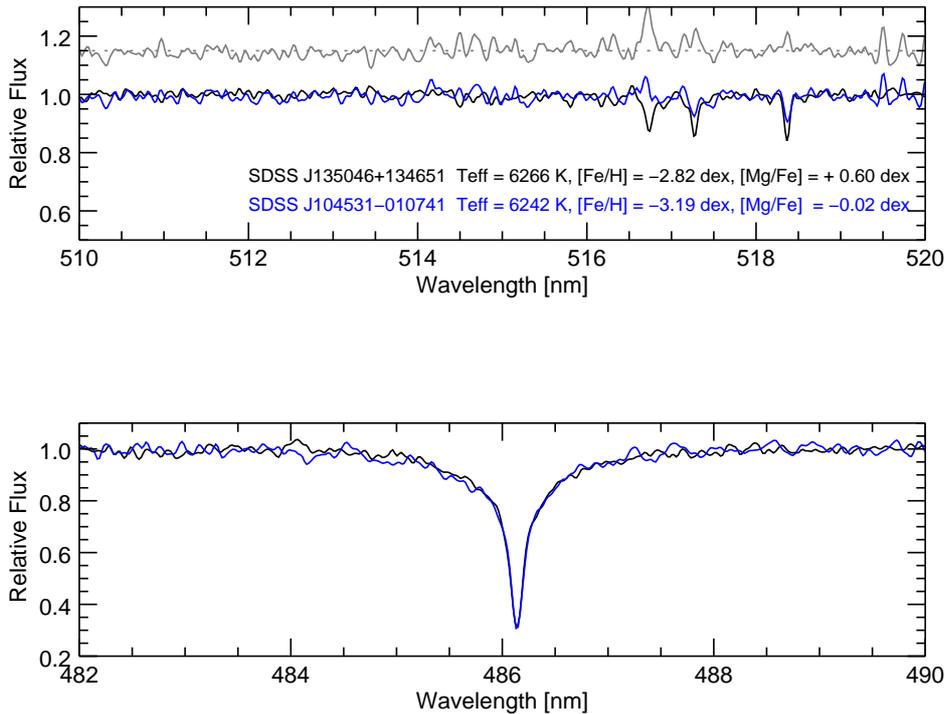}}
   \caption{Comparison of the magnesium triplet  line strength between the two stars with similar stellar parameters SDSS~J135046 and SDSSJ104531.  The flux difference between the two stars is plotted as a grey line on the upper part of the figure. For clarity, this difference has been shifted up by 1.15.  
   The lower part of the plot shows the similar shape of the H {$\beta$}  line for both stars. The difference in [Fe/H] between these two stars is of the order of 0.4 dex. Setting the two stars at the same metallicity [Fe/H] = $-$3.0 dex would increase the difference in the [Mg/Fe] ratio. }
              \label{Mg_comp1}%
    \end{figure}

  \begin{figure}
   \centering
 \resizebox{\hsize}{!}{\includegraphics[clip=true]{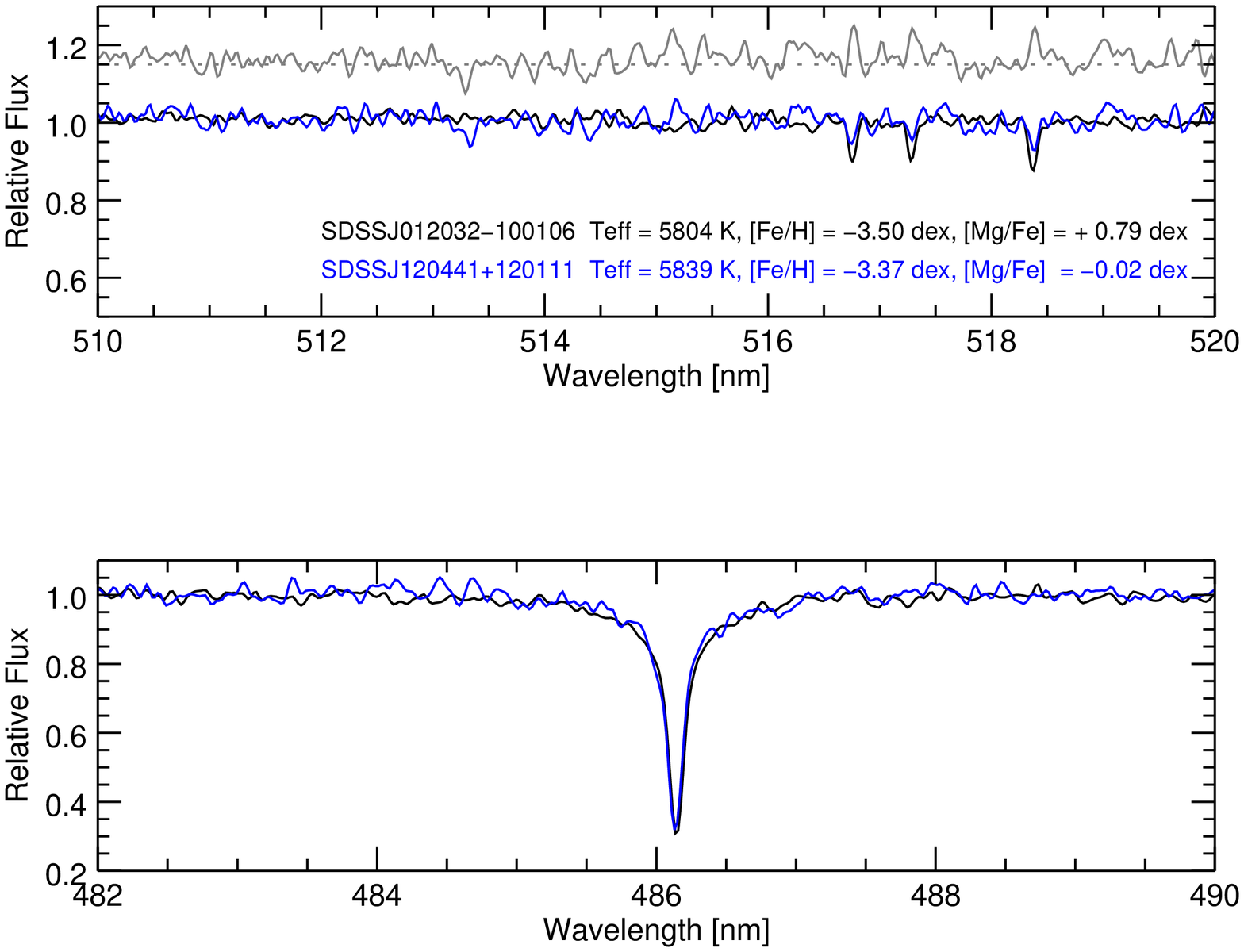}}
   \caption{ Comparison of the magnesium triplet line strength between the two stars with similar stellar parameters SDSSJ012032 and SDSSJ120441. 
   The flux difference between the two stars is plotted as a grey line on the upper part of the figure. For clarity, this difference has been shifted up by 1.15 . 
   The lower part of the plot shows the similar shape of the  H {$\beta$} line for both stars.}
              \label{Mg_comp2}%
    \end{figure}

At a given metallicity, the [Mg/Fe] abundance ratios exhibit a significant scatter, which was already visible in the previous sample of TOPOS results represented as grey
squares in  Fig. \ref{alpha}. To verify the presence of the scatter, we identified "twin stars" in our sample, i.e.  stars with similar atmospheric parameters. We find two couples: the first pair
SDSS~J135046$+$134651  and SDSS~J104531$-$010741 has an effective temperature of $\simeq$ 6200~K and [Fe/H] $\simeq$ $-$3.00~dex and the second pair SDSS~J120441$+$120111 and SDSSJ012032$-$100106,  with a  $T_{\mathrm{eff}}$$\simeq$ 5820~K and [Fe/H]  $\simeq$ $-$3.45dex. 

In Fig. \ref{Mg_comp1} and Fig. \ref{Mg_comp2}, we plot the region where the magnesium triplet (top panel) and H$\beta$ (bottom panel) features form and over$-$plotted the two sets of "twin stars" over one-another. It is clear that neither "twin star" share a similar magnesium abundance, based on the difference in their magnesium line strengths. The difference in  the line strength of the Mg triplet line favours the existence of a real spread  of 
the [Mg/Fe] ratio in stars with [Fe/H]  $\lesssim$ $-$3.00~dex. The same conclusion can be drawn when inspecting the Ca and Si transitions in the same way in these two pairs of stars.

\subsection{Temperature trends}
 \begin{figure}
   \centering
 \resizebox{\hsize}{!}{\includegraphics[clip=true]{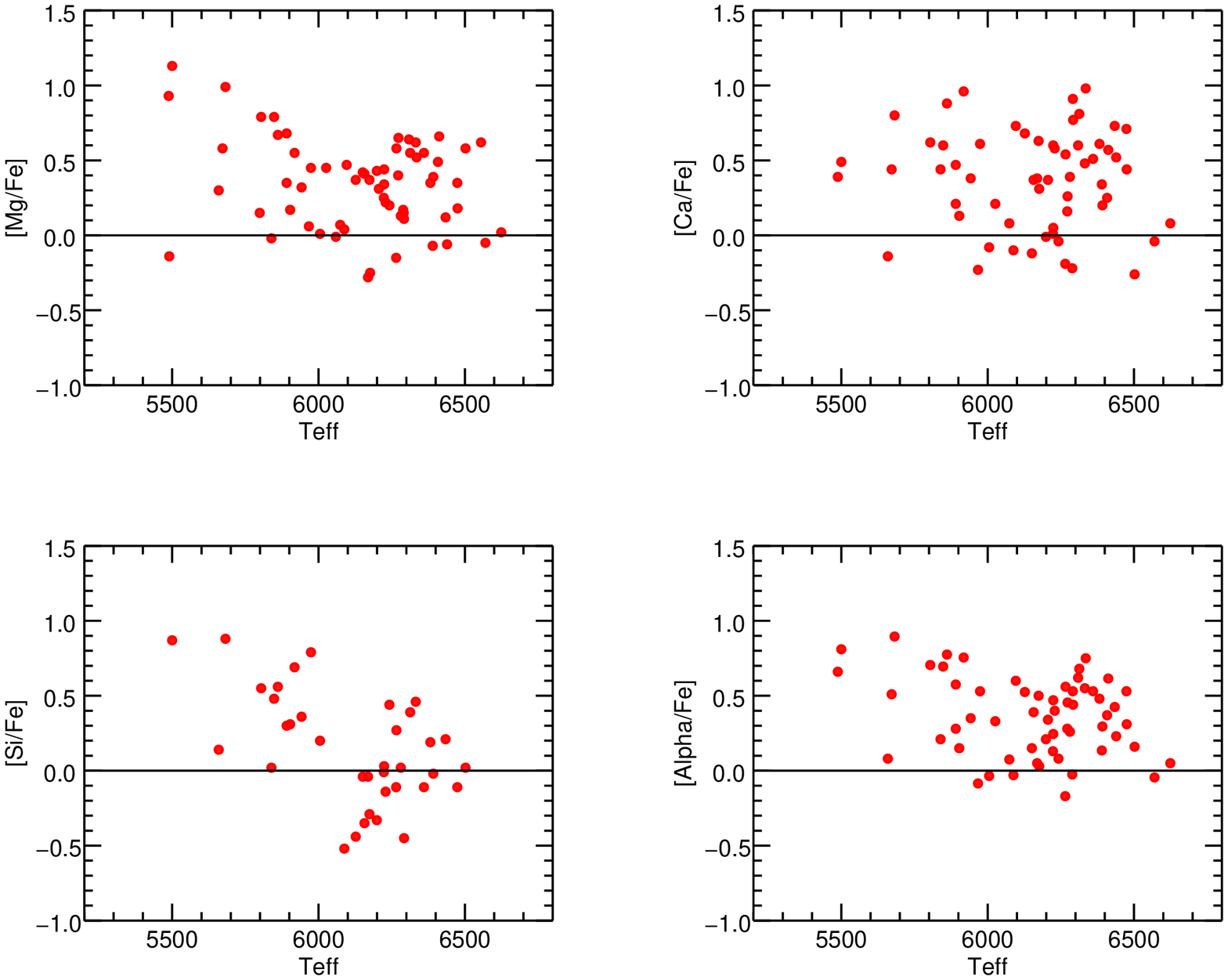}}
   \caption{[Mg/Fe],  [Ca/Fe], [Si/Fe]  and  [$\alpha$/Fe] ratios (evaluated as 0.5$\cdot$([Mg/Fe]+ [Ca/Fe])  as a function of the effective temperature of the star.}
              \label{Teff_trend}%
    \end{figure}

The abundance trends as a function of effective temperature   in a sample of dwarf stars can be used to evaluate the presence of  significant unknown absorption lines in the region  of the transitions we studied.  The strength of this unknown line would change as a function of temperature and would affect the derived abundance. 

In Fig. \ref{Teff_trend}, we show the ratios [Mg/Fe],  [Ca/Fe], [Si/Fe]   and the [$\alpha$/Fe] ratios (evaluated as 0.5 $\times$ ([Mg/Fe]+ [Ca/Fe])  as a function of the effective temperature of the star.  We do not find any  trend of the ratios [Mg/Fe],  [Ca/Fe] and  [$\alpha$/Fe]   or their dispersion as a function of temperature;  we measure a correlation coefficient of the order of 0.05 for the 3 sets of data.     For Si, the results seem to indicate a variation of its abundance  as a function of metallicity, an effect already found by \citet{preston2006} . From their analysis based  on high resolution  high SNR spectra, they conclude that the silicon abundances measured in the 
cooler stars represent the  true abundances of Si, hence an super-solar ratio. 
In the lower right panel of this figure, we  present the mean abundance of magnesium and calcium, in order to minimize 
random errors.
 It is interesting to note a small decrease of the spread of the  [$\alpha$/Fe] ratios at a given temperature compared to the  [Mg/Fe] vs $T_{\mathrm {eff}}$  and  [Ca/Fe] vs $T_{\mathrm {eff}}$.
 However the spread  is still present and is not correlated with temperature.

\subsection{Strontium and barium}

  \begin{figure}
   \centering
 \resizebox{\hsize}{!}{\includegraphics[clip=true]{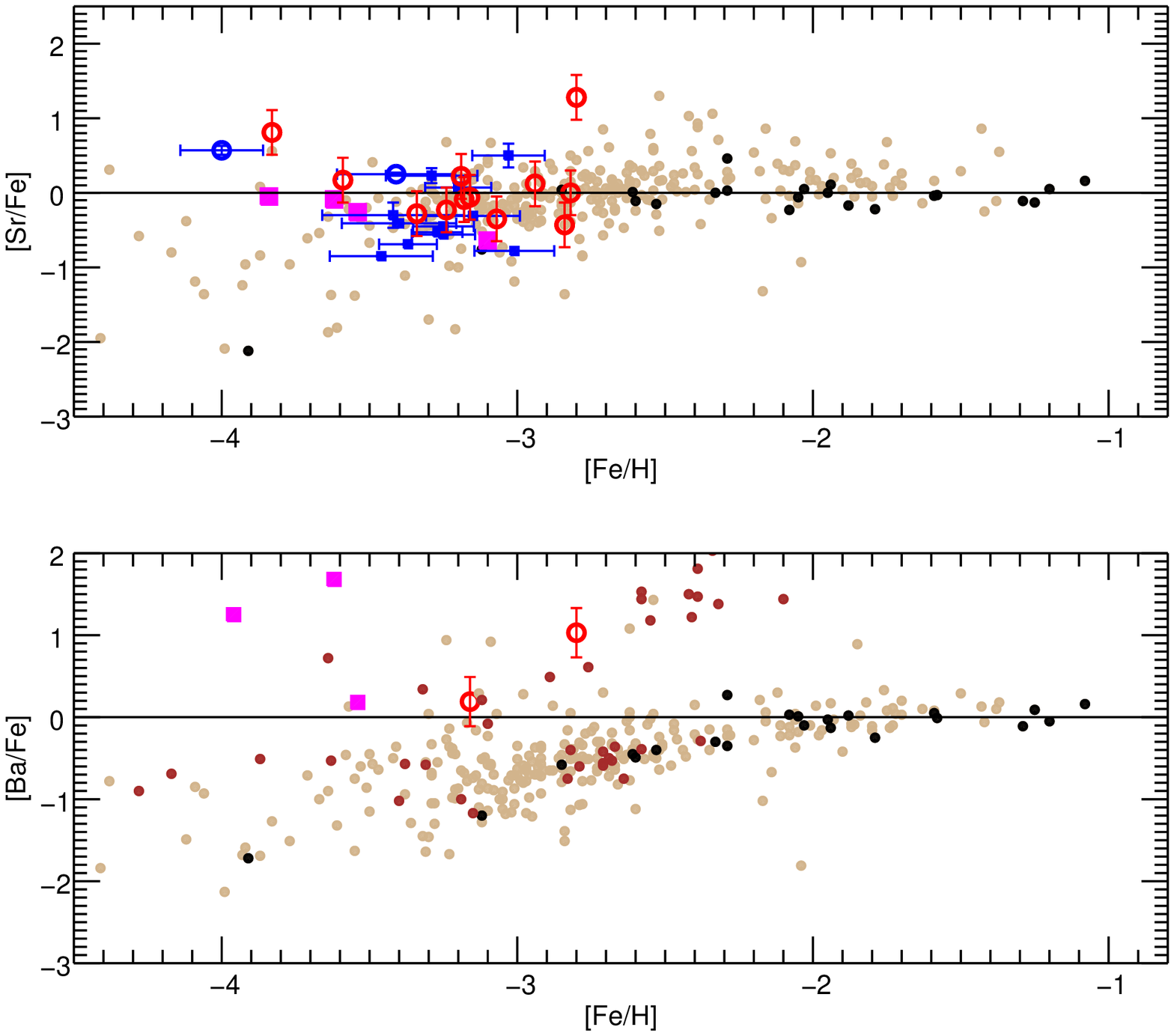}}
   \caption{Neutron$-$capture elements. Red: our results, light brown : \citet{roederer2014}, black: Main$-$sequence stars from \citet{roederer2014}, brown : CEMP stars from \citet{roederer2014}, blue symbols: metal$-$poor dwarf stars \citep{bonifacio2009,  bonifacio2012, bonifacio2018,  caffau2013a,  caffau2013b}. Open circles (resp. squares)  represent the stars which have been observed with UVES (resp. Xshooter).  Pink squares represent dwarf stars analysed by \citep{matsuno2017}. }
              \label{ncapture}%
    \end{figure}

  \begin{figure}
   \centering
 \resizebox{\hsize}{!}{\includegraphics[clip=true]{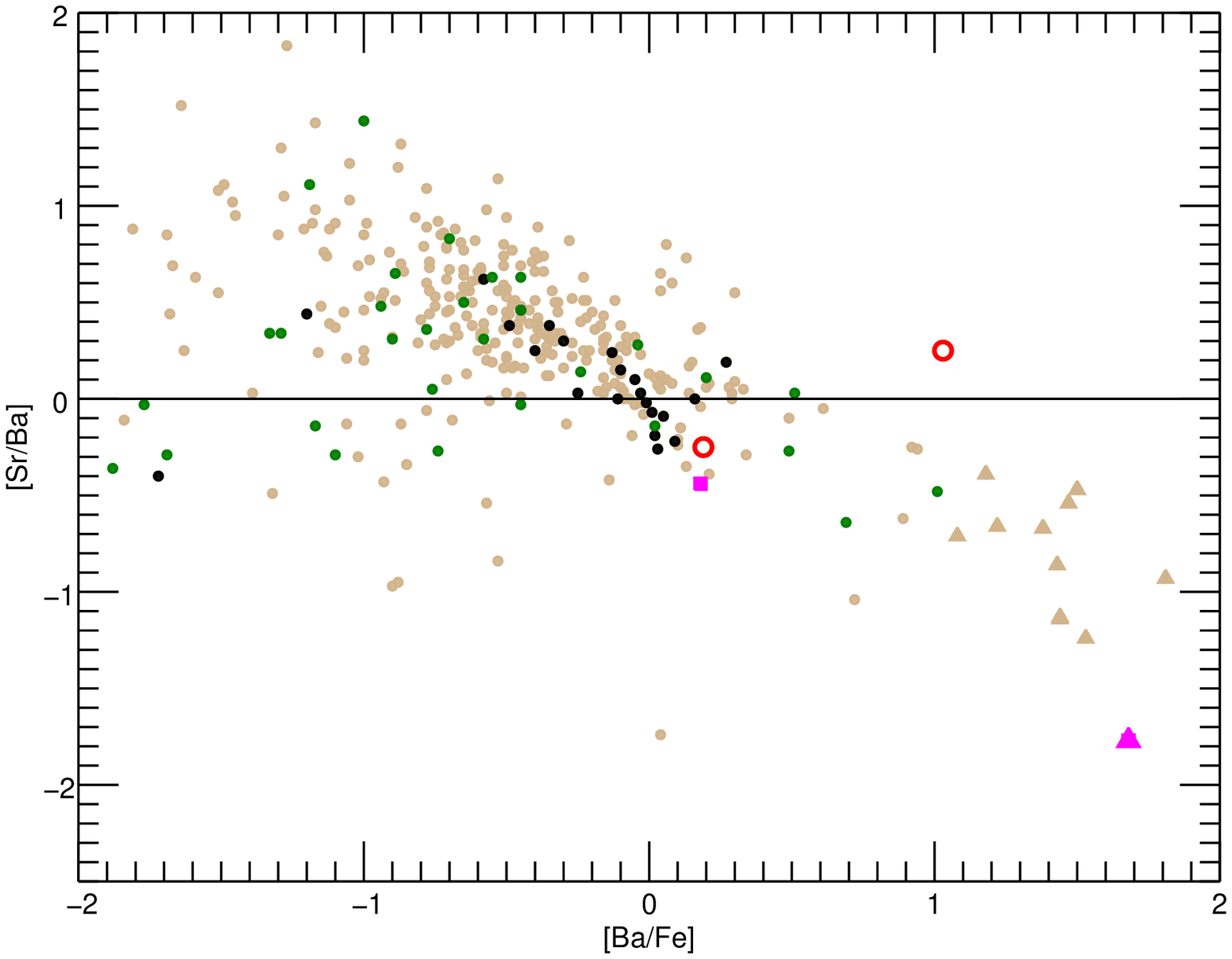}}
   \caption{  [Sr/Ba] vs [Ba/Fe]. Red: our results, light brown:  giant stars \citep{roederer2014},  black: Main$-$sequence stars from \citet{roederer2014}, Pink symbols represent dwarf stars analysed by \citet{matsuno2017}, green: results from \citet{francois2007}, triangles: CEMP stars.}
               \label{ncapture_ratio}%
    \end{figure}

In Fig. \ref{ncapture}, we plot our results for strontium and barium  as red open circles together with the results from   our group  \citep{bonifacio2009,  bonifacio2012, bonifacio2018,  caffau2013a,  caffau2013b}. Blue symbols represent the metal-poor dwarf stars observed by our group \citep{bonifacio2009, bonifacio2012, bonifacio2018,  caffau2013a,  caffau2013b}, the open blue circles denote stars which have been observed with UVES, whereas the blue squares are from X-Shooter spectra. We also add the results of \citet{roederer2014} as light brown circles for evolved stars, and black circles for main$-$sequence and turn$-$off stars. The recent results of \citet{matsuno2017} are represented as pink squares. Their sample has been selected from the Sloan Digital Sky Survey/Sloan Extension for Galactic Understanding and Exploration (SDSS/SEGUE), with follow-up observations with the Subaru telescope. 
In their paper, they selected eight unevolved stars with ${\rm T}_{\rm eff} \le 5500 ~{\rm K}$ and ${\rm [Fe/H]} \le $-$3.5 ~{\rm dex} $.  They could measure the strontium abundance in four of them, and the barium abundance in three of them. The two most metal$-$poor stars of their sample are suspected to be  CEMP (carbon$-$ enhanced$-$metal$-$poor) s-stars, and can hence be considered as different 
from the stars of our sample, which does not contain any CEMP stars.  For strontium, we report a good agreement with the results published in the literature for other turn$-$off stars. 
The situation is more complex for barium.  We find a systematic difference with the trend found by \citet{roederer2014} for their sample of dwarf stars. 
However, this point has to be taken with caution as it is based on a very small sample.  For the majority of the stars of our sample, we could not measure 
the barium abundance. Our results may  simply reflect the limits of detection of the barium abundance in  metal$-$poor turn$-$off stars  from medium resolution and moderate SNR ratios.

The non CEMP s-star from the \citet{matsuno2017} sample are above the solar [Ba/Fe] ratios as well. It is interesting to note that the sample of giant stars of  \citet{roederer2014} includes a significant number of stars with super-solar  [Ba/Fe] ratios. Our stars seem to line up on this upper branch of stars, which have a rather high  [Ba/Fe] ratio.  

 In order to investigate the nature of the stars found in this upper branch with super$-$solar [Ba/Fe] ratios, we identified the CEMP  stars (i.e. with [C/Fe] $>$ +1 dex) of the  sample of \citet{roederer2014} with brown circle symbols in the [Ba/Fe] vs [Fe/H]  plot.  The upper branch is mostly populated with  CEMP  evolved stars  with a moderate [C/Fe] enhancement between 1 and 1.5 dex.  Among them, the CEMP stars with  [Fe/H] $>$ -2.6 dex are  all CEMP$-$s evolved stars leading to the conclusion that this upper branch 
reflects chemical characteristics from the  population of CEMP stars, hence non related with the general galactic chemical evolution.  However, there are several stars in this  upper branch which are not CEMP$-$s nor CEMP stars. This would  be in favor of the existence of a  real bimodal distribution. It is also interesting to note that none of main sequence stars of the sample of \citet{roederer2014} belong to this upper branch. Further studies would be useful to refine the chemical diagnostic of these stars in this upper branch and 
evaluate the reality of a bimodal distribution.

The star SDSS J114424-004658 belonging to our sample shows both a high [Sr/Fe] and [Ba/Fe] ratio (larger than 1 dex).  We also compute an upper limit of Europium for this star and found [Eu/Fe] $<$ 3.10 dex. 
This very high upper limit cannot be used to  demonstrate that this star is a r$-$process enriched star. 
High resolution, high SNR spectra would be necessary to derive  the Europium abundance or at least a useful upper limit. 
This star is very similar to SDSSJ022226.20--031338.0 ([Fe/H]=$-$2.6 dex , [Ba/Fe] $\sim$ 2 dex , [Sr/Ba]$\sim$ 0.3 dex) studied by \citet{caffau2018}.
As SDSS J114424-004658 is not too metal$-$poor  and not too faint (g=17.29), many neutron-capture absorption lines  may be visible in high S/N high resolution spectra.  This  would be hence  a very interesting target  for follow-up observations.

In Fig.  \ref{ncapture_ratio}, we plot the ratio [Sr/Ba] as a function of [Ba/Fe] for the two stars of our sample for which we could measure the abundances of these elements.  
We add the results of \citet{roederer2014} as light brown circles  black circles for main$-$sequence and turn$-$off stars.  We also plot the results for Sr and Ba  from the ESO large programme  "First Stars"  \citep{francois2007}.The two pink squares represent the results of  \citet{matsuno2017}.  The CEMP stars, which  may to be considered as chemically peculiar stars, are represented as triangles. 
 It is interesting to note the increase of the [Sr/Ba] ratio as [Ba/Fe] decreases, as mentioned for example by \citet{spite2018} and \citet{francois2007}.  Merging  the results of \citet{roederer2014} and \citet{matsuno2017}  shows that the decreasing (upper envelope) of the  [Sr/Ba] abundance ratio  as [Ba/Fe] increases  seems to extend  when [Ba/Fe]  becomes super-solar. One of the stars for which we derived both Sr and Ba  (SDSS J114424$-$004658) does not follow this general trend.  
 The star with  [Ba/Fe] larger than 1 dex  and negative [Sr/Ba]  values in this plot is the  CEMP-s star SDSS~J1036+1212 \citep{matsuno2017}, who confirmed the abundances found by  \citet{behara2010}.

\section{Conclusions}

In the context of the TOPoS large programme, we have analysed sixty five metal$-$poor turn$-$off stars using X$-$Shooter spectra increasing by more than a factor of two the number of metal-poor  turn$-$off stars  with detailed  abundances published in the literature.  We measured the abundances of magnesium, calcium and silicon for most of the stars. We were able to derive the abundance of strontium in 12 stars and the abundance of barium in two stars of the sample. 
Both  [Mg/Fe]  and [Si/Fe] seem to show increasing relative abundance as [Fe/H] decreases down to [Fe/H] $\simeq$ $-$4.0 dex.  
However, in the metallicity interval $-$4 to $-$3 dex, there is a significant spread in [$\alpha$/Fe].  
For strontium, we found from solar to  slightly sub$-$solar  [Sr/Fe] ratios,  in agreement with the results published in the literature. 
 Two stars exhibits  a high [Sr/Fe] (around  +1 dex).  The two stars for which we could measure the barium abundance show a super-solar ratio with one 
 of them  (SDSS J114424-004658) with a high [Ba/Fe] ratio of 1.28 dex. This star  has a also a high [Sr/Fe] ratio and hence is an interesting target for follow$-$up observations.  This star is very similar to SDSSJ022226.20--031338.0 ([Fe/H]=$-$2.6 dex , [Ba/Fe] $\sim$ 2 dex , [Sr/Ba]$\sim$ 0.3 dex) studied by \citet{caffau2018}.These super-solar ratios are also found by \citet{matsuno2017} in contrast with the results found by \citet{roederer2014}, who report a systematic 
 sub-solar [Ba/Fe] ratio at the metal-poor end of their sample of turn-off stars. However, it is important to add that  this effect is not found in their giants, for which they found both sub- and super-solar [Ba/Fe] ratios. Their results seem to indicate the presence of two distinct groups of stars  in the metallicity range $-$3.2 ~dex$~ \le$ [Fe/H] $\le$ $-$2.2~dex, one  sample of stars with high [Ba/Fe] and a second sample with low [Ba/Fe].   However, a substantial fraction of the evolved stars found in this upper branch are CEMP$-$s or CEMP stars.  If further studies reveal that this is the case for all the stars in the upper branch,  this would then reflect the peculiar chemical characteristics of this type of stars with no direct impact on the galactic chemical evolution.
 
  It would be very interesting to study the barium abundance in turn-off stars in the metallicity range $-$3.2 ~dex$ \le$ [Fe/H] $\le$ $-$2.2~dex  to  investigate the reality  of a "bimodal" distribution of the [Ba/Fe] ratios as a function of metallicity.

\begin{acknowledgements}
 PF, PB, EC, MS and FS acknowledge support from the Programme National
de Physique Stellaire (PNPS) and the Programme National de Cosmologie
et Galaxies (PNCG) of the Institut National des Sciences de l'Univers of
the CNRS. AJG, NC, AK  and HL were supported by Sonderforschungsbereich SFB 881 "The Milky Way System"  (subprojects A04, A05, A08) of the German Research Foundation (DFG). 
This work has made use of data from the European Space Agency (ESA) mission
{\it Gaia} (\url{https://www.cosmos.esa.int/gaia}), processed by the {\it Gaia}
Data Processing and Analysis Consortium (DPAC,
\url{https://www.cosmos.esa.int/web/gaia/dpac/consortium}). Funding for the DPAC
has been provided by national institutions, in particular the institutions
participating in the {\it Gaia} Multilateral Agreement.
We thank the anonymous referee for the very constructive report.  
\end{acknowledgements}

%
%

\begin{appendix}
\setcounter{table}{1}
\section{Additional Tables}

 \begin{longtable}{l r r c c c c}
  \caption{Stellar parameters } \\
  \hline\hline
   STAR$^a$            &   $\rm T_{eff}$    &   [Fe/H]  &  $\alpha_{2000}^b$& $\delta_{2000}^b$ & G$^b$ & SNR@450 nm\\       
      \endfirsthead
\caption{continued.}\\
\hline\hline
   STAR            &   $\rm T_{eff}$    &   [Fe/H]  &  $\alpha_{2000}$& $\delta_{2000}$ & G & SNR@450 nm  \\       
\hline
\endhead
\hline
SDSS J000411--055027&6174 &--2.96 & 00:04:11.61  & --05:50:28 & 19.0269  & 39\\ 
SDSS J002558--101509&6408 &--3.08 & 00:25:58.60  & --10:15:09 & 17.1013 &  52\\
SDSS J003507--005037&5891 &--2.95 & 00:35:07.72  & --00:50:38 & 17.3268 & 75\\
SDSS J003954--001856&6382 &--2.94 & 00:39:54.66  & --00:18:57 & 17.9868 & 63\\
SDSS J012032--100106&5804 &--3.50 & 01:20:32.63  & --10:01:07 & 16.3543 & 75\\
SDSS J012125--030943&6127 &--2.91 & 01:21:25.11  & --03:09:44 & 18.0162 & 36\\
SDSS J012442--002806&6273 &--2.79 & 01:24:42.11  & --00:28:07 & 18.9395 & 32\\
SDSS J014036+234458&5848 &--3.83 & 01:40:36.22  & +23:44:58 & 15.3520 & 69\\
SDSS J014721+021819&5967 &--3.30 & 01:47:21.86  & +02:18:20 & 17.0625 & 57\\
SDSS J014828+150221&6151 &--3.41 & 01:48:29.01  & +15:02:21 & 17.9115 & 66\\
SDSS J021238+013758&6291 &--3.34 & 02:12:38.49  & +01:37:58 & 17.2226 & 53\\
SDSS J021554+063901&6005 &--2.75 & 02:15:54.33  & +06:39:01 & 18.7562 & 37\\
SDSS J030549+050826&6309 &--2.98 & 03:05:49.64  & +05:08:27 & 18.9920 & 43\\
SDSS J031348+011456&6335 &--3.31 & 03:13:48.15  & +01:14:56 & 19.0425 & 38\\
SDSS J035925--063416&6281 &--3.17 & 03:59:25.57  & --06:34:16 & 18.0274 & 66\\
SDSS J040114--051259&5500 &--3.62 & 04:01:14.73  & --05:12:59 & 18.1712 & 33\\
SDSS J074748+264543&6434 &--3.36 & 07:47:48.61  & +26:45:43 & 17.0241 & 76\\
SDSS J075338+190855&6439 &--2.45 & 07:53:38.62  & +19:08:56 & 17.0073 & 38\\
SDSS J080336+053430&6360 &--2.94 & 08:03:36.58  & +05:34:30 & 16.7814 & 56 \\
SDSS J082506+192753&6390 &--3.07 & 08:25:06.66  & +19:27:53 & 17.6095 & 69\\
SDSS J085232+112331&6206 &--3.45 & 08:52:32.90  & +11:23:31 & 18.0203 & 63\\
SDSS J090533--020843&5974 &--3.44 & 09:05:33.36  & --02:08:45 & 16.4991 & 54\\
SDSS J091913+232738&5490 &--3.25 & 09:19:13.14  & +23:27:38 & 18.3596 & 25\\
SDSS J103402+070116&6224 &--3.58 & 10:34:02.70  & +07:01:17 & 17.2439 & 57\\
SDSS J104531--010741&6242 &--3.19 & 10:45:31.22  & --01:07:42 & 19.0079 & 56\\
SDSS J105002+242109&5682 &--3.93 & 10:50:02.36  & +24:21:10 & 17.6611 & 57\\
SDSS J105231--004008&6199 &--2.72 & 10:52:31.68  & --00:40:09 & 18.2504 &30 \\
SDSS J112031--124638&6502 &--3.27 & 11:20:31.81  & --12:46:39 & 17.0484 & 56\\
SDSS J112211--114809&6157 &--3.00 & 11:22:11.82  & --11:48:09 & 18.5406 & 45\\
SDSS J112750--072711&6474 &--3.34 & 11:27:50.90  & --07:27:12 & 17.7014 &  81\\
SDSS J114424--004658&6412 &--2.80 & 11:44:24.62  & --00:46:59 & 17.0684 &  29\\
SDSS J120441+120111&5839 &--3.37 & 12:04:41.39  & +12:01:11 & 16.0913 &  44\\
SDSS J123055+000546&6223 &--3.24 & 12:30:55.21  & +00:05:47 & 14.4984 & 99\\
SDSS J123404+134411&5659 &--3.59 & 12:34:04.57  & +13:44:11 & 16.4659 & 54 \\
SDSS J124121--021228&5672 &--3.47 & 12:41:21.48  & --02:12:29 & 18.9207 & 72 \\
SDSS J124304--081230&5488 &--3.92 & 12:43:04.16  & --08:12:31 & 17.8240 & 53\\
SDSS J124719--034152&6332 &--4.11 & 12:47:19.46  & --03:41:52 & 18.2498 &  84\\
SDSS J131249+001315&6400 &--2.38 & 13:12:49.61  & +00:13:15 & 16.9237 & 34\\
SDSS J131456--113753&6265 &--3.24 & 13:14:56.86  & +11:37:53 & 17.4045 & 69 \\
SDSS J131948+233436&6074 &--2.61 & 13:19:48.62  & +23:34:36 & 18.6605 & 32\\
SDSS J132112+010256&6395 &--2.49 & 13:21:12.17  & +01:02:56 & 18.9137 & 28 \\
SDSS J132508+222424&6292 &--2.60 & 13:25:08.60  & +22:24:25 & 18.7498 & 28 \\
SDSS J135046+134651&6266 &--2.82 & 13:50:46.74  & +13:46:51 & 18.0776 & 69 \\
SDSS J135331--032930&6224 &--3.18 & 13:53:31.00  & --03:29:30 & 16.5132 & 83\\
SDSS J140007+191236&6570 &--3.19 & 14:00:07.59  & +19:12:36 & 19.0487 & 61\\
SDSS J141249+013206&5799 &--2.94 & 14:12:49.07  & +01:32:07 & 17.6992 & 30\\
SDSS J150702+005152&6555 &--3.51 & 15:07:02.01  & +00:51:53 & 18.5433 & 50 \\
SDSS J153747+281404&6272 &-3.39 & 15:37:47.78  & +28:14:05 & 18.0039 & 54 \\
SDSS J154746+242953&5903 &-3.16 & 15:47:46.50  & +24:29:53 & 17.8342 & 51\\
SDSS J155159+253900&6059 &-3.08 & 15:51:59.34  & +25:39:01 & 17.0147 & 35\\
SDSS J155751+190306&6176 &-2.94 & 15:57:51.77  & +19:03:06 & 17.5649 & 35 \\
SDSS J172552+274116&6624 &-2.91 & 17:25:52.21  & +27:41:17 & 19.1727 & 33\\
SDSS J173358+274952&6088 &-3.03 & 17:33:58.00  & +27:49:52 & 18.9242 & 59\\
SDSS J200513-104503&6289 &-3.41 & 20:05:13.50  & -10:45:03 & 16.6015 & 55 \\
SDSS J214633-003910&6475 &-3.07 & 21:46:33.18  & -00:39:10 & 17.8769 & 77\\
SDSS J215023+031928&6026 &-2.84 & 21:50:23.52  & +03:19:28 & 18.5872 &  34\\
SDSS J215805+091417&5942 &-3.41 & 21:58:05.90  & +09:14:17 & 18.0081 & 64\\
SDSS J220121+010055&6392 &-3.03 & 22:01:21.77  & +01:00:55 & 18.3999 & 66 \\
SDSS J220728+055658&6096 &-3.26 & 22:07:28.09  & +05:56:59 & 18.1022 & 40\\
SDSS J222130+000617&5891 &-3.14 & 22:21:30.23  & +00:06:17 & 19.1787 & 33 \\
SDSS J225429+062728&6169 &-3.01 & 22:54:29.63  & +06:27:28 & 18.4909 & 34 \\
SDSS J230243-094346&5861 &-3.71 & 23:02:43.34  & -09:43:46 & 18.7481 &  56\\
SDSS J231031+031847&6229 &-2.91 & 23:10:31.86  & +03:18:48 & 16.7470 & 54\\
SDSS J231755+004537&5918 &-3.54 & 23:17:55.56  & +00:45:38 & 18.0750 & 56\\
SDSS J235210+140140&6313 &-3.54 & 23:52:10.24  & +14:01:40 & 17.9775 & 74\\

 \\ \hline
\multicolumn{6}{l}{$^a$ The name of the star is based on the SDSS DR12 coordinates.}\\
\multicolumn{6}{l}{$^b$ From Gaia DR2.}\\
         \label{stellar_parameters}\\
   \end{longtable}

 \begin{longtable}{l r r r r r r r r r r r r  }
      \caption{Abundance ratios [X/Fe] for the sample of stars. The number of lines used to compute the abundances of Mg, Si, Ca , Sr and Ba are given respectively in columns 5,7,9,11 and 13.  } \\
 Object     &	 [Fe/H] &[C/Fe] &  [Mg/Fe] & n &  [Si/Fe] & n & [Ca/Fe] & n & [Sr/Fe] & n & [Ba/Fe] & n \\ 	
    \label{star_abund}
      \endfirsthead
             
\caption{continued.}\\
\hline\hline
 Object     &	 [Fe/H] & [C/Fe] &  [Mg/Fe] & n &  [Si/Fe] & n & [Ca/Fe]& n & [Sr/Fe]& n  & [Ba/Fe]& n \\ 	
\hline
\endhead
\hline
 \hline
SDSS J000411$-$055027 &    $-$2.96 & < 1.20	&   0.37 & 3 & $-$0.29   & 1 & 0.63    & 1 &  ----   &   &  ----  &    \\ 
SDSS J002558$-$101509 &    $-$3.08 & < 1.52	&   0.49 & 3 &  ----     &   & 0.25    & 1 &  ----   &   &  ----  &    \\
SDSS J003507$-$005037 &    $-$2.95 & < 0.69	&   0.68 & 4 &   0.30    & 1 &   0.47  & 3 &  ----   &  &  ----  &    \\
SDSS J003954$-$001856 &    $-$2.94 & < 1.08	&   0.35 & 3 &   0.19    & 1 &   0.61  & 3 &  ----   &  &  ----  &    \\
SDSS J012032$-$100106 &    $-$3.50 & < 0.94	&   0.79 & 3  &   0.55   & 1 &   0.62  & 2 &  ----   &   &  ----  &    \\
SDSS J012125$-$030943 &    $-$2.91 & < 0.85	&   0.37 & 2 &  $-$0.44  & 1 &   0.68  & 1 &  ----   &   &  ----  &    \\
SDSS J012442$-$002806 &    $-$2.79 & < 1.03	&   0.65 & 3 &     ----  &   &   0.26  & 1 &  ----   &   &  ----  &    \\
SDSS J014036$+$234458 &    $-$3.83 & < 1.27	&   0.79 & 3 &  0.48     & 1 & 0.60    & 1 &  0.81   & 2 &  ----  &    \\
SDSS J014721$+$021819 &    $-$3.30 & < 1.24	&   0.06 & 3 &     ----  &   & $-$0.23 & 2 &  ----   &   &  ----  &    \\
SDSS J014828$+$150221 &    $-$3.41 & < 0.95	&   0.42 & 3 & $-$0.04   & 1 & $-$0.12 & 1 &  ----   &   &  ----  &   \\
SDSS J021238$+$013758 &    $-$3.34 & < 1.58	&   0.15 & 2 &    -----  &   &    0.91 & 2 &  ----   &   &  ----  &    \\
SDSS J021554$+$063901 &    $-$2.75 & < 0.49	&   0.01 & 2 &    0.20   & 1 & $-$0.08 & 2 &  ----   &   &  ----  &    \\
SDSS J030549$+$050826 &    $-$2.98 & < 1.12	&   0.64 & 2 &    ----   &   & 0.60    & 2 &  ----   &   &  ----  &    \\
SDSS J031348$+$011456 &    $-$3.31 & < 1.85	&   0.52 & 2 &  ----     &   & 0.98    & 1 &  ----   &   &  ----  &    \\
SDSS J035925$-$063416 &    $-$3.17 & < 1.41	&   0.13 & 3 &    0.02   & 1 & 0.39    & 2 &  ----   &   &  ----  &    \\
SDSS J040114$-$051259 &    $-$3.62 & < 0.86	&   1.13 & 3 &  0.87     & 1 & 0.49    & 2 &  ----   &   &  ----  &    \\ 
SDSS J074748$+$264543 &    $-$3.36 & < 1.3 	&   0.12 & 2 &    0.21   & 1 & 0.73    & 2 &  ----   &   &  ----  &    \\
SDSS J075338$+$190855 &    $-$2.45 & < 1.39	&  -0.06 & 3 &    ----   &   & 0.52    & 1 &  ----   &   &  ----  &    \\
SDSS J080336$+$053430 &    $-$2.94 & < 1.18	&   0.55 & 3 &   $-$0.11 & 1 & 0.51    & 2 &  0.12   & 2 &  ---   &    \\
SDSS J082506$+$192753 &    $-$3.07 & < 1.31	&  -0.07 & 3 &    ----   &   & 0.34    & 2 &  ----   &   &  ----  &    \\
SDSS J085232$+$112331 &    $-$3.45 & < 1.39	&   0.31 & 3 &    ----   &   & 0.37    & 2 &  ----   &   &  ----  &    \\
SDSS J090533$-$020843 &    $-$3.44 & < 1.78	&   0.45 & 3 &    0.79   & 1 & 0.61    & 2 &  ----   &   &  ----  &    \\
SDSS J091913$+$232738 &    $-$3.25 & < 0.49	&  -0.14 & 3 &  ----     &   &  ----   &  &  ----   &   &  ----  &    \\
SDSS J103402$+$070116 &    $-$3.58 & < 1.52	&   0.44 & 3 &   -----   &   & 0.05    & 2 &  ----   &   &  ----  &    \\
SDSS J104531$-$010741 &    $-$3.19 & < 0.83	&   0.20 & 2 &  0.44     & 1 & $-$0.04 & 2 &  0.22   & 2 &  ----  &    \\
SDSS J105002$+$242109 &    $-$3.93 & < 1.17	&   1.29 & 3 &   0.88    & 1 & 0.00    & 2 &  ----   &   &  ----  &    \\ 
SDSS J105231$-$004008 &    $-$2.72 & < 1.16	&   0.43 & 2 & $-$0.33   & 1 & $-$0.01 & 2 &  ----   &   &  ----  &    \\
SDSS J112031$-$124638 &    $-$3.27 & < 1.71	&   0.58 & 3 &  0.02     & 1 & $-$0.26 & 2 &  ----   &   &  ----  &    \\
SDSS J112211$-$114809 &    $-$3.00 & < 0.94	        &   0.41 & 3 &  $-$0.35  & 1 & 0.37    & 2 &  ----   &   &  ----  &    \\
SDSS J112750$-$072711 &    $-$3.34 & < 1.88	        &   0.35 & 3 & $-$0.11   & 1 & 0.71    & 1 & $-$0.28 & 1 &   ---- &    \\ 
SDSS J114424$-$004658 &    $-$2.80 & < 1.54	&   0.66 & 3 &   ----    &   & 0.57    & 2 &  1.28   & 2 &  1.03  & 1  \\
SDSS J120441$+$120111 &    $-$3.37 & < 0.81	&  -0.02 & 3 &  0.02     & 1 & 0.44    & 2 &  ----   &   &  ----  &    \\
SDSS J123055$+$000546 &    $-$3.24 & < 1.08	&   0.25 & 3 &  $-$0.02  & 1 & 0.01    & 2 & $-$0.23 & 2 & ----   &   \\ 
SDSS J123404$+$134411 &    $-$3.59 & < 0.63	&   0.30 & 3 &  0.14     & 1 &  ----   &   &  0.17   & 1 &  ----  &    \\ 
SDSS J124121$-$021228 &    $-$3.47 & < 0.61	&   0.58 & 2 &  ----     &   & 0.44    & 2 &  ----   &   &  ----  &    \\ 
SDSS J124304$-$081230 &    $-$3.92 & < 0.66	&   0.93 & 3 &   ----    &   & 0.39    & 2 &  ----   &   &  ----  &    \\
SDSS J124719$-$034152 &    $-$4.11 & < 1.75	&   0.62 & 3 &  0.46     & 1 & 0.48    & 1 &  ----   &   &  ----  &    \\
SDSS J131249$+$001315 &    $-$2.38 & < 1.52	&   ---- &   &   ----    &   &  ----   &   &  ----   &   &  ----  &    \\ 
SDSS J131456$-$113753 &    $-$3.24 & < 1.38	&  -0.15 & 3 &  $-$0.11  & 1 & $-$0.19 & 2  &  ----   &   &  ----  &    \\ 
SDSS J131948$+$233436 &    $-$2.61 & < 0.85     &   0.07 & 3 &  ----     &   & 0.08    & 2 &  ----   &   &  ----  &    \\
SDSS J132112$+$010256 &    $-$2.49 & < 1.73	&   ---- &   &   ----    &   & ----    &   &  ----   &   &  ----  &    \\
SDSS J132508$+$222424 &    $-$2.60 & < 1.24	&   0.11 & 3 & $-$0.45   & 1 & 0.77    & 3 &  ----   &   &  ----  &    \\
SDSS J135046$+$134651 &    $-$2.82 & < 1.06	&   0.58 & 3 &  0.27     & 1 & 0.54    & 3 &  0.00   & 2 &  ----- &     \\
SDSS J135331$-$032930 &    $-$3.18 & < 1.12	&   0.34 & 3 &  0.03     & 1 & 0.60    & 1 & $-$0.09 & 2 & -----  &    \\ 
SDSS J140007$+$191236 &    $-$3.19 & < 1.93	&  -0.05 & 2 &  ----     &   & -0.04   & 2 &  ----   &   &  ----  &    \\ 
SDSS J141249$+$013206 &    $-$2.94 & < 0.68	&   0.15 & 3 &  ----     &   & 0.66    & 1 &  ----   &   &  ----  &    \\ 
SDSS J150702$+$005152 &    $-$3.51 & < 1.95	&   0.62 & 2 &  ----     &   & ----    &   &  ----   &   &  ----  &    \\
SDSS J153747$+$281404 &    $-$3.39 & < 1.83	&   0.40 & 3 &  ----     &   & 0.16    & 2 &  ----   &   &  ----  &    \\ 
SDSS J154746$+$242953 &    $-$3.16 & < 0.90	&   0.17 & 3 &  0.31     & 1 & 0.13    & 2 & $-$0.06 & 1 & 0.19   & 1  \\
SDSS J155159$+$253900 &    $-$3.08 & < 1.02	&  -0.01 & 3 &  ----     &   & ----    &   &  ----   &   &  ----  &    \\
SDSS J155751$+$190306 &    $-$2.94 & < 1.08	&  -0.25 & 3 &  ----     &   & 0.31    & 2 &  ----   &   &  ----  &    \\
SDSS J172552$+$274116 &    $-$2.91 & < 1.65	&   0.02 & 3 &  ----     &   & 0.08    & 2 &  ----   &   &  ----  &    \\
SDSS J173358$+$274952 &    $-$3.03 & < 0.87	&   0.04 & 3 & $-$0.52   & 1 & $-$0.10 & 2 &  ----   &   &  ----  &    \\
SDSS J200513$-$104503 &    $-$3.41 & < 1.65	&   0.17 & 3 &  ----     &   & $-$0.22 & 2 &  ----   &   &  ----  &    \\
SDSS J214633$-$003910 &    $-$3.07 & < 1.31	&   0.18 & 3 &   ----    &   & 0.44    & 3 & $-$0.35 & 2 & ----   &   \\
SDSS J215023$+$031928 &    $-$2.84 & < 0.78	&   0.45 & 3 &   ----    &   & 0.21    & 2 & $-$0.43 & 2 &  ----  &   \\
SDSS J215805$+$091417 &    $-$3.41 & < 1.05	&   0.32 & 3 &  0.36     & 1 & 0.38    & 2 &  ----   &   &  ----  &    \\
SDSS J220728$+$055658 &    $-$3.26 & < 1.50	&   0.47 & 3 &  ----     &   & 0.73    & 2 &  ----   &   &  ----  &    \\ 
SDSS J220121$+$010055 &    $-$3.03 & < 1.27	&   0.39 & 3 & -0.02     & 1 & 0.20    & 2 &  ----   &   &  ----  &    \\
SDSS J222130$+$000617 &    $-$3.14 & < 1.28	&   0.35 & 3 &   ----    &   & 0.21    & 1 &  ----   &   &  ----  &    \\
SDSS J225429$+$062728 &    $-$3.01 & < 1.25	&  -0.28 & 3 & $-$0.04   & 1 & 0.38    & 2 &  ----   &   &  ----  &    \\
SDSS J231031$+$031847 &    $-$2.91 & < 1.05	&   0.22 & 3 & $-$0.14   & 1 & 0.58    & 2 &  ----   &   &  ----  &    \\ 
SDSS J231755$+$004537 &    $-$3.54 & < 1.48	&   0.55 & 3 &  0.69     & 1 & 0.96    & 2 &  ----   &   &  ----  &    \\ 
SDSS J230243$-$094346 &    $-$3.71 & < 1.15	&   0.67 & 3 &  0.56     & 1 & 0.88    & 2 &  ----   &   &  ----  &    \\ 
SDSS J235210$+$140140 &    $-$3.54 & < 1.78	&   0.55 & 2 &  0.39     & 1 & 0.81    & 2 &  ----   &   &  ----  &    \\ 
            \noalign{\smallskip}     
            \hline		     
         \end{longtable}

\end{appendix}
\end{document}